\documentclass[amsmath,amssymb,nofootinbib,preprint]{revtex4}

\usepackage{latexsym,comment}
\usepackage{amssymb}
\usepackage{amsfonts,verbatim}
\usepackage{amsmath,color}
\usepackage[dvips]{graphicx}
\usepackage{bm}

\def\ber{\begin{eqnarray}}
\def\eer{\end{eqnarray}}
\def\beq{\begin{equation}}
\def\eeq{\end{equation}}

\newcommand\vep{\varepsilon}

\begin{document}

\title{Light bending in $f(T)$ gravity}

\author{Matteo Luca Ruggiero}
\email{matteo.ruggiero@polito.it}
\affiliation{Dipartimento di Fisica - Universit\`a  di Torino, Via Pietro Giuria 1, Torino, Italy, \\ DISAT - Politecnico di Torino, Corso Duca degli Abruzzi 24,  Torino, Italy\\
 INFN, Sezione di Torino, Via Pietro Giuria 1, Torino, Italy}

\date{\today}

\begin{abstract}
In the framework of $f(T)$ gravity, we focus on a weak-field  and spherically symmetric solution  for the Lagrangian $f(T)=T+\alpha T^{2}$, where $\alpha$ is a small constant which parameterizes the departure from General Relativity. In particular, we study  the propagation of light  and obtain the correction to the general relativistic bending angle. Moreover, we discuss the impact of this correction on some gravitational lensing observables, and evaluate the possibility of constraining the theory parameter $\alpha$ by means of observations.  {In particular, on taking into account the astrometric accuracy  in the Solar System, we obtain that $|\alpha| \leq  1.85 \times 10^{5}\, \mathrm{m^{2}}$; this bound is looser than those deriving from the analysis of Solar System dynamics, e.g. $|\alpha| \leq 5 \times 10^{-1}\, \mathrm{m^{2}}$ \cite{Iorio:2015rla},  $|\alpha| \leq 1.8 \times 10^{4}\, \mathrm{m^{2}}$ \cite{Iorio12} or $|\alpha| \leq 1.2 \times 10^{2}\, \mathrm{m^{2}}$ \cite{Xie:2013vua}. However we suggest that,  since the effect only depends on the impact parameter,  better constraints could be obtained by studying light bending from planetary objects.}  
\end{abstract}

\maketitle

\section{Introduction}\label{sec:intro}

The so-called $f(T)$  theories \cite{Cai:2015emx} are a generalization  of Teleparallel Gravity (TEGR) \cite{pereira,Aldrovandi:2003xu,Maluf:2013gaa}; the latter is a theory of gravity based on a Riemann-Cartan space-time, endowed with the non symmetric Weitzenb\"ock connection which, unlike the Levi-Civita connection of GR, gives rise to torsion but  is curvature-free. In TEGR  torsion plays the role of curvature, while the tetrad, instead of the metric,  plays the role of the dynamical field; the field equations are obtained from a Lagrangian containing the torsion scalar $T$. Actually, even if TEGR has a different geometric structure with respect to General Relativity (GR), the two theories have the same dynamics: in other words  every solution of GR is also solution of TEGR. In   the $f(T)$  theories  the  Lagrangian is an  analytic function of the torsion scalar $T$: these theories  generalize  TEGR and  are not equivalent to GR  \cite{Ferraro:2008ey, Fiorini:2009ux}. For this reason they have been considered as potential  candidates to solve the issue of cosmic acceleration \cite{cardone12,Myrzakulov:2010vz,Nashed:2014lva,Yang:2010hw,bengo,kazu11,Karami:2013rda,sari11,cai11,capoz11,Bamba:2013jqa,Camera:2013bwa,Nashed:2015pda}.  The additional degrees of freedom of $f(T)$  gravity are related to the fact that the equations of motion are not invariant under local Lorentz transformations \cite{Li:2010cg}: for this reason it is  important to suitably choose a tetrad that does not constrain a priori the functional form of the Lagrangian \cite{tamanini12}. In recent papers \cite{Krssak:2015lba,Krssak:2015oua} the problem of the violation of local Lorentz invariance has been analyzed with emphasis on the role of the spin connection, and it has been showed that it is possible to obtain a fully covariant reformulation of $f(T)$ gravity.

Spherically symmetric solutions are important for $f(T)$ gravity because they can be used to constrain these theories in the Solar System. To this end, a weak-field solution  for a Lagrangian in the form $f(T)=T+\alpha T^{2}$ (where $\alpha$ is a small constant which parameterizes the departure from GR) has been obtained by Iorio\&Saridakis \cite{Iorio12}: this solution has been used to constrain the $\alpha$ parameter in the Solar System \cite{Xie:2013vua}. In a subsequent paper Ruggiero\& Radicella \cite{Ruggiero:2015oka} have obtained a new solution for a  Lagrangian in the general form $f(T)=T+\alpha T^{n}$, with $|n| \neq 1$: we refer to the solution obtained for $n=2$  as the RR solution.  A preliminary analysis of the impact of the RR solution on the Solar System dynamics has been carried out in \cite{Iorio:2015rla}. 

In this paper we focus on the propagation of light  in the RR space-time and study the corrections to the GR bending angle, due to the non linearity of Lagrangian. In particular, we exploit the  general approach for light bending and gravitational lensing in arbitrary spherically symmetric space-times introduced in \cite{Keeton:2005jd}, and study the lensing observables in the RR space-time. Then, we use these results to evaluate the possibility of constraining the theory parameter $\alpha$.

This work is organized as follows: in Section \ref{sec:fT-gravity} we briefly review  the foundations of $f(T)$ theories, in order to obtain the RR space-time; then, in Section \ref{sec:lensing}, we study  light propagation and, in Section \ref{sec:disc}, we evaluate the impact  on  the observations. Conclusions are  in Section \ref{sec:conc}.

\section{Field equations and spherically symmetric solutions}\label{sec:fT-gravity}

In this Section, in order to make the paper self-consistent, 
 we show how the RR solution is obtained in the framework of $f(T)$ gravity. To begin with, we recall that in this theory the tetrad plays the role of the dynamical field instead of the metric:  given a coordinate basis, the components $e^a_\mu$ of the tetrad are related to the metric tensor $g_{\mu\nu}$ by $g_{\mu \nu}(x) = \eta_{a b} e^a_\mu(x) e^b_\nu(x)$, with  $\eta_{a b} = \text{diag}(1,-1,-1,-1)$. Notice that, here and henceforth,  latin indexes refer to the tangent space, while greek indexes label coordinates on the manifold, and we use units such that $G=c=1$ (if not otherwise specified). We get the field equations by varying the action 
\begin{equation}
{\cal{S}} = \frac{1}{16 \pi } \int{ f(T)\, e \, d^4x} + {\cal{S}}_M \ ,
\label{eq:action}
\end{equation}
with respect to the tetrad, where   $e = \text{det} \  e^a_\mu = \sqrt{-\text{det}(g_{\mu \nu})}$ and ${\cal{S}}_M$ is the action for the matter fields; $f$ is a differentiable function of the torsion scalar $T$: in particular, if $f(T)=T$, the action is the same as in TEGR, and the theory is equivalent to GR.  The torsion tensor is defined by 
\beq
T^\lambda_{\ \mu \nu} = e^\lambda_a \left( \partial_\nu e^a_\mu - \partial_\mu e^a_\nu \right ), \  \label{eq:deftorsiont}
\eeq
and the contorsion tensor by
\beq
S^\rho_{\ \mu \nu} = \frac{1}{4} \left ( T^{\rho}_{\ \ \mu \nu} - T_{\mu \nu}^{\ \ \rho}+T_{\nu \mu}^{\ \ \rho} \right ) +
\frac{1}{2} \delta^\rho_\mu T_{\sigma \nu}^{\ \ \sigma} - \frac{1}{2} \delta^\rho_\nu T_{\sigma \mu}^{\ \ \sigma}.  \label{eq:defcontorsion}
\eeq
Eventually,  the torsion scalar is 
\beq
T = S^\rho_{\ \mu \nu} T_\rho ^{\ \mu \nu}.  \label{eq:deftorsions}
\eeq

The variation of the action (\ref{eq:action}) with respect to the tetrad field gives the field equations
\begin{widetext}
\beq
e^{-1}\partial_\mu(e\  e_a^{\ \rho}   S_{\rho}^{\ \mu\nu})f_T+e_{a}^{\ \lambda} S_{\rho}^{\ \nu\mu} T^{\rho}_{\ \mu\lambda} f_T
+  e_a^{\ \rho}  S_{\rho}^{\ \mu\nu}\partial_\mu (T) f_{TT}+\frac{1}{4}e_a^{\nu} f = 4\pi  e_a^{\ \mu} {\cal{T}}_\mu^\nu,
\label{eq: fieldeqs}
\eeq
\end{widetext}
in terms of the matter-energy tensor ${\cal{T}}^\nu_\mu$;  the subscripts $T$, here and henceforth, denote differentiation with respect to $T$.

We are interested in spherically symmetric solutions that can be used to describe the gravitational field of a point-like source. To this end, we write the space-time metric in the form 
\begin{equation}
ds^2=e^{A(r)}dt^2-e^{B(r)}dr^2-r^2 d\Omega^2 \ , \label{metric}
\end{equation}
where $d\Omega^{2}= d\theta^2+ \sin^2 \theta d\phi^2$ is the space metric on the unit sphere. 

Actually, vacuum spherically symmetric solutions are always in the form of the Schwarzschild-de Sitter metric if the torsion scalar is constant, i.e. $dT/dr=0$, as it is shown in \cite{tamanini12}. So, we can get new solutions  if we assume that $dT/dr\neq 0$. To this end, we use the non diagonal tetrad introduced in \cite{tamanini12}
$$
e_\mu^a=\left( \begin{array}{cccc}
e^{A/2}         &   0                                             &   0                                            &    0         \\
0                 &e^{B/2} \sin{\theta}\cos{\phi}   & e^{B/2} \sin{\theta}\sin{\phi}&  e^{B/2} \cos{\theta}\\
0                 &-r \cos{\theta}\cos{\phi}   & -r  \cos{\theta}\sin{\phi}&  r \sin{\theta}\\
0                 &r \sin{\theta}\sin{\phi}   & -r  \sin{\theta}\cos{\phi}&  0\\
 \end{array} \right) 
$$
to obtain the field equations. 

Indeed, a diagonal tetrad that gives back the metric in eq. (\ref{metric}) is not a good choice since the equations of motion for such a choice would constrain a priori the form of the Lagrangian. This is related to the lack of the local Lorentz invariance of $f(T)$ gravity:  tetrads connected by local Lorentz transformations lead to the same metric - i.e. the same causal structure - but different equations of motions, thus physically inequivalent solutions (see \cite{tamanini12}). 

We obtain the following field equations in vacuum (see e.g. \cite{Ruggiero:2015oka}):
\begin{widetext} 
\begin{eqnarray}
&&\frac{f(T)}{4}-f_T\frac{e^{-B (r)}}{4r^2}\left(2-2e^{B(r)}+r^2 e^{B(r)} T -2r B'(r)\right)+\nonumber\\
&&-f_{TT} \frac{T'(r) e^{-B(r)}}{r}\left(1+e^{B(r)/2}\right)=0 \label{00eq1}\\
&&-\frac{f(T)}{4}+f_T\frac{e^{-B (r)}}{4r^2}\left(2-2e^{B(r)}+r^2 e^{B(r)} T -2r A'(r)\right)=0 \label{11eq1}\\
&&f_T\left[-4+4e^{B(r)}-2r A'(r)-2r B'(r)+r^2A'(r)^2-r^2A'(r)B'(r)+2r^{2}A''(r)\right]+\nonumber\\
&&+2rf_{TT}T'\left(2+2e^{B(r)/2}+rA'(r)\right)=0 \label{eq31}
\end{eqnarray}
\end{widetext}
The  torsion scalar turns out to be
\begin{equation}\label{torsionscalar}
T=\frac{2e^{-B(r)}(1+e^{B(r)/2})}{r^{2}}\left[1+e^{B(r)/2}+r A'(r)\right].
\end{equation}

We are looking for  weak-field solutions, so we can safely suppose that these solutions are perturbations of a flat background Minkowski space-time. As a consequence, we write $e^{A(r)}=1+A(r), \quad e^{B(r)}=1+B(r)$ for the metric coefficients. Furthermore, in solving the field equations (\ref{00eq1})-(\ref{eq31}) we confine ourselves to linear perturbations, and consider $f(T)$ in the form  $f(T)=T+\alpha T^2$. This Lagrangian is interesting since it can be considered as the first approximation of a power series expression of an arbitrary Lagrangian:   $\alpha$ is expected to be a small constant, parameterizing the departure of these theories from GR (or from TEGR, which is the same).  The solution is
\beq
A(r)=-{\frac {{ 2M}}{r}}-32\,{\frac {\alpha}{{r}^{2}}} \label{eq:sola}
\eeq
\beq
B(r) =+{\frac {{2M}}{
r}}+96\,{\frac {\alpha}{{r}^{2}}} \ , \label{eq:solb}
\eeq

so that we can write the RR metric in the form
\beq
ds^{2}=-\left(1-{\frac {{ 2M}}{r}}-32\,{\frac {\alpha}{{r}^{2}}} \right)dt^{2}+\left(1+\frac {2M}{
r}+96\,{\frac {\alpha}{{r}^{2}}} \right)dr^{2}+r^{2}d\Omega^2 \ . \label{eq:RRmetric}
\eeq
In the above solution there is a correction, proportional to $\alpha$, to the gravitational field (in weak-field approximation) of a point-like source of mass $M$; we stress that the above solution is approximated up to linear order both in $M/r$ and $\alpha/r^{2}$.

\section{Light Propagation }\label{sec:lensing}

In order to study the propagation of light in the metric (\ref{eq:RRmetric}), we follow the general approach described in \cite{Keeton:2005jd}  for arbitrary static spherically symmetric space-times. In particular, we focus on the bending of light rays in the RR metric and, moreover, we describe the corrections to some lensing observables due to non linearity of the Lagrangian.

We start from a metric written in the form:
\beq
ds^{2}=-a(r)dt^{2}+b(r)dr^{2}+r^{2}d\Omega^{2} \ ,\label{eq:metricasimm}
\eeq
and we express the coefficients in power series:
\begin{eqnarray}
a(r) & = & 1+2a_{1}\phi+2a_{2} \phi^{2}+2a_{3} \phi^{3}+... \label{eq:Apower} \ , \\
b(r) & = & 1-2b_{1}\phi+4b_{2} \phi^{2}-8b_{3} \phi^{3}+... \label{eq:Bpower}  \ , 
\end{eqnarray}
where $\displaystyle \phi=-\frac{M}{r}$ is the Newtonian potential. We confine ourselves to the second order approximation to make a comparison with the RR metric. On setting $\displaystyle \sigma=\frac{\alpha}{M^{2}}$,  from (\ref{eq:RRmetric}),  (\ref{eq:Apower}) and (\ref{eq:Bpower}), we get 
\beq
\left.\begin{array}{cc}a_{1}=1 & a_{2}=-16 \sigma \ , \\ b_{1}=1 & b_{2}=24 \sigma \ . \end{array}\right. \label{eq:aibi}
\eeq

The bending angle $\hat \varepsilon$ can be written as a series expansion which expresses the correction to the weak-field bending angle of GR. If we define the gravitational radius\footnote{In physical units: $\displaystyle r_{S}=\frac{GM}{c^{2}}$.} $\displaystyle r_{g}=M$, the GR bending due to a mass $M$ is $\displaystyle\hat \varepsilon_{GR}=\frac{4r_{g}}{b}$, where $b$ is the impact parameter. The bending angle in the metric (\ref{eq:metricasimm}) can be written as\footnote{We confine ourselves to the first correction to the GR value, however in \cite{Keeton:2005jd} the bending angle is obtained up to the third order.}:
\beq 
  \hat \varepsilon = A_1 \left(\frac{r_{g}}{b}\right) \ + \ A_2 \left(\frac{r_{g}}{b}\right)^2\ + \  O\left(\frac{r_{g}}{b}\right)^{3} \ , \label{eq:hatalpha}
\eeq
where the coefficients $A_1,A_{2}$ are independent of $M/b$.  In terms of the coefficients $a_{1},b_{1},a_{2},b_{2}$, we have $A_{1}=2\left(a_{1}+b_{1} \right)$, $A_{2}=\left(2a_{1}^{2}-a_{2}+a_{1}b_{1}-\frac{b_{1}^{2}}{4}+b_{2} \right)\pi$. In particular, on taking into account  (\ref{eq:aibi}), we get
\beq
A_{1}=4, \quad A_{2}=\left(40\sigma+\frac{11}{4} \right)\pi \ . \label{eq:A1A2}
\eeq

If we explicitly write the expression (\ref{eq:hatalpha}) of the bending angle, we get $\displaystyle \hat \varepsilon= \frac{4M}{b}+ \left( 40 \sigma + \frac{11}{4} \right)\pi \frac{M^{2}}{b^{2}}$; since $\displaystyle \sigma = \frac{\alpha}{M^{2}}$, we eventually have $\displaystyle \hat \varepsilon=\frac{4M}{b}+  40 \pi \frac{\alpha}{b^{2}} + \frac{11\pi }{4}  \frac{M^{2}}{b^{2}}$. However, since the RR metric (\ref{eq:RRmetric}) has been obtained up to linear order in $M/r$, we must neglect the contribution proportional to $\displaystyle \frac{M^{2}}{b^{2}}$:  consequently, here and henceforth, we set  $A_{2}= 40\sigma \pi$. 

Hence, the bending angle turns out to be
\beq
\hat \varepsilon=  \frac{4M}{b} +\frac{40\alpha \pi}{b^{2}} \ ,  \label{eq:bensingsigma}
\eeq
or
\beq
\hat \varepsilon=\hat \varepsilon_{GR} \left( 1 +\frac{10 \alpha \pi}{Mb} \right) \ . \label{eq:bensingsigmabis}
\eeq

The above result is in agreement with \cite{Bozza:2015haa}, where the bending angle is calculated in spherically symmetric  metrics falling as $\displaystyle \frac{1}{r^{q}}$.

\begin{figure}[top]
\centering
\includegraphics[scale=.55]{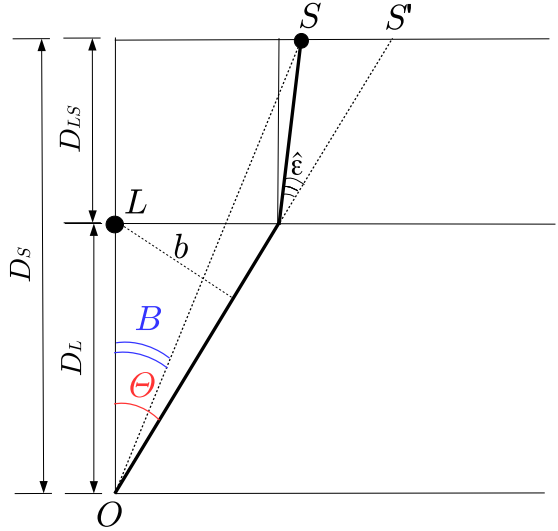}
\caption{Lens Geometry.} \label{fig:lens}
\end{figure}

Let us recall the basic theory of gravitational lensing. The geometry of lensing is described in Figure \ref{fig:lens}. The angular positions of the source and the image are $B$ and $\Theta$, while $\hat \varepsilon$ is the bending angle; $D_{L},D_{S}, D_{LS}$ are the observer-lens, observer-source, lens-source distances. The observer in $O$ sees the image of the source, located at $S$, as if it were in $S'$. The lens is located at the point $L$. In the so-called thin lens approximation  the light paths are approximated by straight lines. Let $b$ be  the impact parameter, which is a constant of motion  of light propagation: it is the perpendicular distance (relative to inertial observers at infinity) from the center of the lens to the asymptotic tangent line to the light ray trajectory to the observer; from the figure we see that $b=D_{L} \sin \Theta$. The following \textit{lens equation} can be obtained by means of elementary geometric relations: 
\beq
D_{S} \tan B = D_{S} \tan \Theta-D_{LS} \left[\tan \Theta-\tan \left(\Theta-\hat \varepsilon \right) \right] \ .\label{eq:lenseq}
\eeq
The latter equation allows to obtain the angular position of the image as a function of the angular position of the source and the  bending angle. We assume that the lens is static and spherically symmetric, and that both the observer and the source are in the asymptotically flat zone of space-time. Furthermore, we suppose that the light rays propagates outside the gravitational radius $\displaystyle r_{g}$ of the source: in other words, if $r_{0}$ is the distance of closest approach, we suppose that $r_{0} \gg r_{g}$.

If the angles are small (weak lens approximation), the lens equation  (\ref{eq:lenseq}) can be written as $\displaystyle D_{S} B=D_{S} \Theta- D_{LS} \hat \varepsilon$. On using the  GR expression of the bending angle $\displaystyle \hat \varepsilon_{GR}=\frac{4M}{b}$ and  the impact parameter $b=D_{L}  \Theta$, the lens equation becomes
\beq
D_{S} B =D_{S} \Theta- \frac{D_{LS}}{D_{L}}\frac{4 M}{\Theta} \ . \label{eq:lensapprox}
\eeq

The  solution of this equation for $B=0$, which describes the alignment of observer, lens and source, defines the so-called \textit{Einstein angle} $\theta_{E}$:
\beq
\theta_{E} = \sqrt{\frac{4M D_{LS}}{D_{L}D_{S}}} \ ,\label{eq:Eiangle}
\eeq
which is a characteristic angular scale; correspondingly, it is possible to define the \textit{Einstein radius} $R_{E}=D_{L} \theta_{E}$, which is a characteristic length scale. We scale all angular positions with $\theta_{E}$:
\beq
\beta=\frac{B}{\theta_{E}}, \, \, \theta=\frac{\Theta}{\theta_{E}} \ ,
\eeq
Moreover we set: $\displaystyle   \epsilon=\frac{\Theta_{M}}{\theta_{E}}$, where $\Theta_{M}=\tan^{-1}(M/D_{L})$ is the angle subtended by the gravitational radius of the lens. The parameter  $\epsilon$ is used to  expand the lensing observables in power series: from  the lens equation (\ref{eq:lenseq}) and postulating that $\hat \varepsilon$ is  in the form (\ref{eq:hatalpha}), the image position can be written as
\beq
\theta=\theta_{0}+\theta_{1}\epsilon+ O(\epsilon^{2}) \ , \label{eq:powerexpansionepsilon}
\eeq
where $\theta_{0}$ (i.e. the image position in the weak-field deflection limit) is the solution of 
\beq
0=-\beta+\theta_{0}-\frac{1}{\theta_{0}} \ . \label{eq:lens0}
\eeq
Accordingly, we get the images position 
\beq 
  \theta_0^\pm = \frac{1}{2}\left(\sqrt{ 4 + \beta^2} \pm |\beta| \right) , \label{eq:imagepos0}
\eeq
where $\theta_{0}^{+}$ is the positive parity image, lying on the same side of the lens as the source ($\beta>0$), while he negative parity image $\theta_{0}^{-}$ lies on the opposite site of the lens from the source ($\beta<0$). 

The second order term turns out to be (see e.g. \cite{Keeton:2005jd}):
\beq 
  \theta_1 = \frac{A_2}{A_1+4\theta_0^2}\ . \label{eq:theta1}
\eeq

Summarizing, up to first order in $\epsilon$; the image position can be written as
\beq
\theta=\theta_{0}+ \frac{A_2}{A_1+4\theta_0^2} \epsilon \ . \label{eq:thetatot1}
\eeq

Hence, in our case, taking into account the values (\ref{eq:A1A2}) of $A_{1}, A_{2}$, for the RR metric we have
\beq
\theta=\theta_{0}+ \frac{40 \sigma \pi }{4+4\theta_0^2} \epsilon =\theta_{0}+ \frac{10 \sigma \pi }{1+\theta_0^2} \epsilon \ . \label{eq:thetatot1fT}
\eeq

Since the actual angular positions are given by $\Theta=\theta \theta_{E}$, the correction can be written as $\Theta_{1}=\theta_{1} \theta _{E} \epsilon$, which for small angles can be estimated by  $\displaystyle \Theta_{1} \simeq \theta_{1} \frac{M}{D_{L}} \simeq 10 \pi \sigma \frac{M}{D_{L}}$.

In this formalism, it is possibile to obtain the (signed) magnification $\mu$ of an image at angular position $\Theta$, which has the general expression $\displaystyle \mu(\Theta) = \left[\frac{\sin B (\Theta)}{\sin \Theta} \     \frac{d \ B(\Theta)}{d \Theta} \right]^{-1}$. The series expansion in $\epsilon$ is written as
\beq
\mu= \mu_{0}+\mu_{1} \epsilon + O(\epsilon^{2}) \ , \label{eq:mu1}
\eeq
where
\beq
\mu_0 = \frac{16\theta_0^4}{16\theta_0^4-A_1^2}, \quad \quad  \mu_1 = - \frac{16 A_2 \theta_0^3}{(A_1+4\theta_0^2)^3}\  . \label{eq:mu2}
\eeq

Then, on taking into account the values  (\ref{eq:A1A2}) of $A_{1}, A_{2}$, for the RR metric, we  obtain

\beq
\mu_0 = \frac{\theta_0^4}{\theta_0^4-1}, \quad \quad  \mu_1 = - \frac{ 10 \sigma \pi \theta_0^3}{(1+\theta_0^2)^3}\ . \label{eq:mu3}
\eeq

Remember that $\mu>0$ for the positive-parity image $\theta^{+}$, while $\mu<0$ for the negative-parity image $\theta^{-}$. Consequently, we see that the sign of $\sigma=\alpha/M^{2}$ influences the magnification: if $\alpha>0$, $\mu_{1}$ is negative, so the positive-parity image is fainter, while the negative-parity image is brighter. In principle, this could provide an observational test for the theory parameter $\alpha$. The total magnification is not modified up to first order in $\epsilon$, while the second order term is  proportional to $A_{2}^{2}$, hence it is null in our approximation.

It is possible to work out the time delay, i.e. the difference between the actual light travel time and the travel time light would take if the lens were not present: we have
\beq
  \frac{\tau}{\tau_E} =
    \frac{1}{2}
    \left[ a_1 + \beta^2 - \theta_0^2 - \frac{a_1+b_1}{2} \ln \left(
    \frac{D_L\,\theta_0^2\,\theta_E^2}{ 4\,D_{LS}} \right) \right]
\ + \ \frac{\pi}{16\,\theta_0}
    \Bigl( 8 a_1^2 - 4 a_2 + 4 a_1 b_1 - b_1^2 + 4 b_2 \Bigr)\,\epsilon
\ + \ O({\epsilon^{2}}) \ ,   \label{eq:delay111}
\eeq
which in our  case becomes
\beq
 \frac{\tau}{\tau_E}= \frac 1 2 \left[1+\beta^{2}-\theta_{0}-\ln \left(
    \frac{D_L\,\theta_0^2\,\theta_E^2}{ 4\,D_{LS}} \right)  \right]+\frac{1}{\theta_0} \left(10 \pi\sigma \right)\epsilon \ . \label{eq:timedelay1}
\eeq
where\footnote{In physical units $\displaystyle \tau_{E}=4\frac{GM}{c^{3}}$.} $\tau_{E}=4M$ is a natural time scale of the system. Moreover, it is possible to obtain the differential time delay between the positive and negative parity images:
\beq \label{eq-PPN-dtau}
  \Delta\tau = \Delta\tau_0 \ + \ \vep\,\Delta\tau_1 \ + O(\epsilon^{2}) ,
\eeq
where
\begin{eqnarray}
  \Delta\tau_0 &=& \tau_E
    \left[ \frac{(\theta_0^{-})^{-2} - (\theta_0^{+})^{-2}}{2}
    - \frac{a_1+b_1}{2}\,\ln\left(\frac{\theta_0^{-}}{\theta_0^{+}}\right)
    \right] , \\
  \Delta\tau_1 &=& \tau_E\ \frac{\pi}{16}
    \Bigl( 8 a_1^2 - 4 a_2 + 4 a_1 b_1 - b_1^2 + 4 b_2 \Bigr)
    \frac{(\theta_0^{+}-\theta_0^{-})}{\theta_0^{+} \theta_0^{-}}\ .
\end{eqnarray}

In particular, for the RR metric on using the values (\ref{eq:aibi}) of $a_{1},b_{1},a_{2},b_{2}$ we have
\begin{eqnarray}
  \Delta\tau_0 &=& \tau_E
    \left[ \frac{(\theta_0^{-})^{-2} - (\theta_0^{+})^{-2}}{2}
    -\ln\left(\frac{\theta_0^{-}}{\theta_0^{+}}\right)
    \right] , \\
  \Delta\tau_1 &=& \tau_E
    \Bigl( 10 \pi \sigma \Bigr)
    \frac{(\theta_0^{+}-\theta_0^{-})}{\theta_0^{+} \theta_0^{-}}\ . \label{eq:tau1}
\end{eqnarray}

The order of magnitude of the first order correction to the time delay is $\Delta \tau_{1} \epsilon \simeq \tau_{E} 10 \pi \sigma \epsilon$.

\section{Discussion}\label{sec:disc}

Let us discuss the possibility of constraining the theory parameter $\alpha$ appearing in the Lagrangian with the study of light propagation. In previous works, the Lagrangian $f(T)=T+\alpha T^{2}$ and, in particular, the RR solution (\ref{eq:RRmetric}), has been constrained by means of the dynamics of Solar System bodies:    the upper bound for $\alpha$ is $5 \times 10^{-1} \, \mathrm{m^{2}}$ in \cite{Iorio:2015rla}.  Actually, the same Lagrangian has been constrained on the basis of astronomical observations and Solar System experiments in \cite{Xie:2013vua}, starting from the spherically symmetric solution obtained in \cite{Iorio12} which, however, has been obtained by using a diagonal tetrad: as we have seen above (see also the discussion in \cite{Ruggiero:2015oka}), this fact limits the self-consistency of the solution.

From Eq. (\ref{eq:bensingsigmabis}) we may write the deviation in the bending angle from the GR value in the form
\beq
\Delta_{\varepsilon}=|\hat \varepsilon- \hat \varepsilon_{GR}|= \frac{40\pi |\alpha|}{b^{2}} \ .\label{eq:deltaepsilon1} 
\eeq

The accuracy available in astrometry, thanks to the Very Large Baseline Array (VLBA), is of the order of $(10  \simeq 100) \, \mathrm{\mu as}$ (see e.g. \cite{Fomalont:2009zg,Fomalont:2003pd, Brunthaler:2005qs}). Accordingly, we may set the following upper bound on $|\alpha|$:
\beq
|\alpha| \leq \left (\frac{\Delta_{\varepsilon}}{10 \, \mathrm{\mu as}} \right) \left(\frac{b^{2}}{R^{2}_{\odot}} \right) 1.85 \times 10^{5}\, \mathrm{m^{2}}\ , \label{eq:boundalpha}
\eeq
where we have used as reference for the impact parameter the radius of the Sun $R_{\odot}$. We emphasize that the correction due to the non linearity of the Lagrangian is independent of the mass. Furthermore, using this upper bound, we can easily check that $\displaystyle \frac{\alpha}{R_{\odot}^{2}} \simeq 3.8 \times 10^{-13}$ is small enough and can safely be considered as a perturbation of the flat Minkowski space-time, which we have assumed to obtain the RR solution.

The upper bound in Eq. (\ref{eq:boundalpha}) is looser than the one already available, deriving from the analysis of Solar System dynamics. However, since the $\alpha$ contribution to the bending angle only depends on the impact parameter, tighter constraints could be obtained with objects smaller than the Sun: for instance, the contribution from Jupiter would give (with the same astrometric accuracy) a bound smaller of two orders of magnitude. Perhaps, this could be of some interest for the ongoing Gaia mission \cite{Crosta:2005ch,Gaia}.

Eventually, let us discuss the impact of $\alpha$ on  lensing observables. As we have seen above in Eq. (\ref{eq:thetatot1fT}) the image position is influenced by $\alpha$. Let us consider, for instance, the case of the supermassive black hole at the center of the Milky Way. If we take $D_{L}=7600$ pc, and consider a source near the black hole with $D_{LS}=10$ pc, we have $D_{S} \simeq D_{L}$, the Einstein radius is $\theta_{E} = 7.3 \times 10^{-2}\, \mathrm{\mu as}$ and, hence, the lensing scale is $R_{E} = 2.6 \times 10^{-3}$ pc, which is much greater than the one that we have considered above for light bending by the Sun, or Jupiter.  Furthermore, according to the discussion above, the order of magnitude of the correction to the image position is $\displaystyle \Theta_{1} \simeq \ 10 \pi \sigma \frac{M}{D_{L}}=10 \pi \frac{\alpha}{M D_{L}}$: it is evident that this correction decreases with the mass of the lens and it is not effective for supermassive objects. Similarly, we can estimate the first order correction to the differential time delay (\ref{eq:tau1}). It turns out to be $\displaystyle \Delta \tau_{1} \epsilon \simeq \tau_{E} 10 \pi \sigma \epsilon \simeq \frac{10 \pi \alpha}{D_{L} \theta_{E}}$: again, we see that this correction decreases with the mass of the lens ($\theta_{E}$ depends on the mass of the lens).

 In summary, lensing observables in the case of supermassive objects would give poor constraints on the theory parameter $\alpha$.

\section{Conclusions}\label{sec:conc}

In the framework of $f(T)$ gravity,  we have focused on a spherically symmetric solution for the Lagrangian $f(T)=T+\alpha T^{2}$, that can be considered as the first approximation of a power series expression of an arbitrary Lagrangian. The $\alpha$ parameter measures the deviation from General Relativity or, which is the same, from Teleparallel Gravity. In previous works, this parameter has   been constrained analyzing Solar System dynamics; here, by exploiting a general formalism that applies to spherically symmetric space-times, we have studied the deflection of light. We have obtained the correction to the GR bending angle and, moreover, using the lensing formalism, we have derived the modifications of the images position, the time delay, and the magnification. 

These corrections are negligibly small for supermassive objects: this is ultimately related to the fact that the modification to the bending angle does not depend on the mass of the lens. However, the sign of $\alpha$ influences the magnification: images with different parity undergo opposite effects;  in principle, this could constrain the sign of $\alpha$. 

Eventually, we have showed that the study of light deflection with the VLBA accuracy  could provide constraints on $\alpha$ that are looser than those already available. Indeed, since the effect only depends on the impact parameter, we have suggested  that the study of gravitational bending from planetary objects  could give better constraints: perhaps this could be of some relevance for the astrometric missions, such as Gaia. 

{As a final remark, we remember that   $f(T)$ theories  generalize  TEGR just as  $f(R)$ theories do for GR: indeed,  in $f(R)$ gravity (which is based on the Levi-Civita connection of GR) the gravitational Lagrangian depends on a function $f$ of the curvature scalar $R$
(see \cite{Sotiriou:2008rp,Capozziello:2007ec,DeFelice:2010aj} and references therein): when $f(R)=R$ the action reduces to the usual Einstein-Hilbert action, and Einstein's theory is obtained. Both theories, as discussed in the above references, have been considered to explain observations at large cosmological scales, so it is in general expected that their impact at the Solar System scale is very small, as we have discussed in this work. We  have investigated lensing  in $f(R)$ gravity elsewhere \cite{Ruggiero:2007jr}, and we have showed that  the non linearity of the Lagrangian could be relevant for  distant galaxies or clusters of galaxies acting as lenses. However, in order to make a possible comparison between $f(T)$ and $f(R)$ gravity on the basis of lensing observables, it is necessary to generalize  the  simple point-like lens model that we have used in this paper. }

\end{document}